\documentclass[%
    reprint,
 superscriptaddress,
 nofootinbib,
 amsmath,amssymb,
 aps,
 prx,
]{revtex4-2}
\usepackage{graphicx}
\usepackage{dcolumn}
\usepackage{bm}
\usepackage{braket}
\usepackage[hidelinks,colorlinks=true]{hyperref}
\usepackage{nicefrac}
\usepackage{stmaryrd}
\usepackage{multirow, makecell} 
\usepackage[normalem]{ulem}

\usepackage{tikz}
\usetikzlibrary{quantikz2}

\newcommand{\im}{\mathrm{i}}
\newcommand{\e}{\mathrm{e}}

\begin{document}

\title{Quantum-HPC hybrid computation of biomolecular excited-state energies}

\def\quantinuumLondon{Quantinuum Ltd., Partnership House, Carlisle Place, London SW1P 1BX, United Kingdom}
\def\quantinuumTokyo{Quantinuum K.K., Otemachi Financial City Grand Cube 3F, 1-9-2 Otemachi, Chiyoda-ku, Tokyo, Japan}
\def\quantinuumColorado{Quantinuum, LLC, 303 South Technology Court, Broomfield, Colorado 80021, USA}
\def\rikeniTHEMS{RIKEN Center for Interdisciplinary Theoretical and Mathematical Sciences (iTHEMS), RIKEN, Wako, Saitama 351-0198, Japan}
\def\quantinuumCambridge{Quantinuum Ltd., Terrington House, 13-15 Hills Road, Cambridge CB2 1NL, UK}
\def\rikenRCCS{RIKEN Center for Computational Science, RIKEN, Kobe, Hyogo 650-0047, Japan}

\author{Kentaro Yamamoto}
\email{kentaro.yamamoto@quantinuum.com}
\affiliation{\quantinuumTokyo}

\author{Riku Masui}
\affiliation{\quantinuumTokyo}

\author{Takahito Nakajima}
\affiliation{\rikenRCCS}

\author{Miwako Tsuji}
\affiliation{\rikenRCCS}

\author{Mitsuhisa Sato}
\affiliation{\rikenRCCS}

\author{Peter Schow}
\affiliation{\quantinuumColorado}

\author{Lukas Heidemann}
\affiliation{\quantinuumCambridge}

\author{Matthew Burke}
\affiliation{\quantinuumCambridge}

\author{Philipp Seitz}
\affiliation{\quantinuumCambridge}

\author{Oliver J. Backhouse}
\affiliation{\quantinuumCambridge}

\author{Juan W. Pedersen}
\affiliation{\quantinuumTokyo}

\author{John Children}
\affiliation{\quantinuumCambridge}

\author{Craig Holliman}
\affiliation{\quantinuumTokyo}

\author{Nathan Lysne}
\affiliation{\quantinuumTokyo}

\author{Daichi Okuno}
\thanks{His current affiliation is Yaqumo Inc., 1F Yusen Bldg 2-3-2 Marunouchi, Chiyoda-ku Tokyo 100-0005, Japan }
\affiliation{\quantinuumTokyo}

\author{Seyon Sivarajah}
\affiliation{\quantinuumCambridge}

\author{David Mu\~{n}oz Ramo}
\affiliation{\quantinuumCambridge}

\author{Alex Chernoguzov}
\affiliation{\quantinuumColorado}

\author{Ross Duncan}
\affiliation{\quantinuumCambridge}

\date{\today}
             
\begin{abstract}
    We develop a workflow within the ONIOM framework and demonstrate it on the hybrid computing system consisting of the supercomputer Fugaku and the Quantinuum Reimei trapped-ion quantum computer. This hybrid platform extends the layered approach for biomolecular chemical reactions to accurately treat the active site, such as a protein, and the large and often weakly correlated molecular environment. Our result marks a significant milestone in enabling scalable and accurate simulation of complex biomolecular reactions
\end{abstract}

\maketitle


Biomolecular chemical reactions~\cite{Migliore2014-xs,Yano2014-vt} are fundamental to various biological processes, including enzymatic reactions~\cite{Villa2001-ti,Warshel1976-ua} and photoreception~\cite{Kiser2014-op}. Accurate simulation of these reactions is crucial for understanding their mechanisms and for applications in biotechnology, such as drug design. However, simulating biomolecular reactions poses significant challenges due to the complex interplay between the active site, where the reaction occurs, and the surrounding environment, which can be large in molecular size.

\begin{figure}
    \centering
    \includegraphics[width=0.99\hsize]{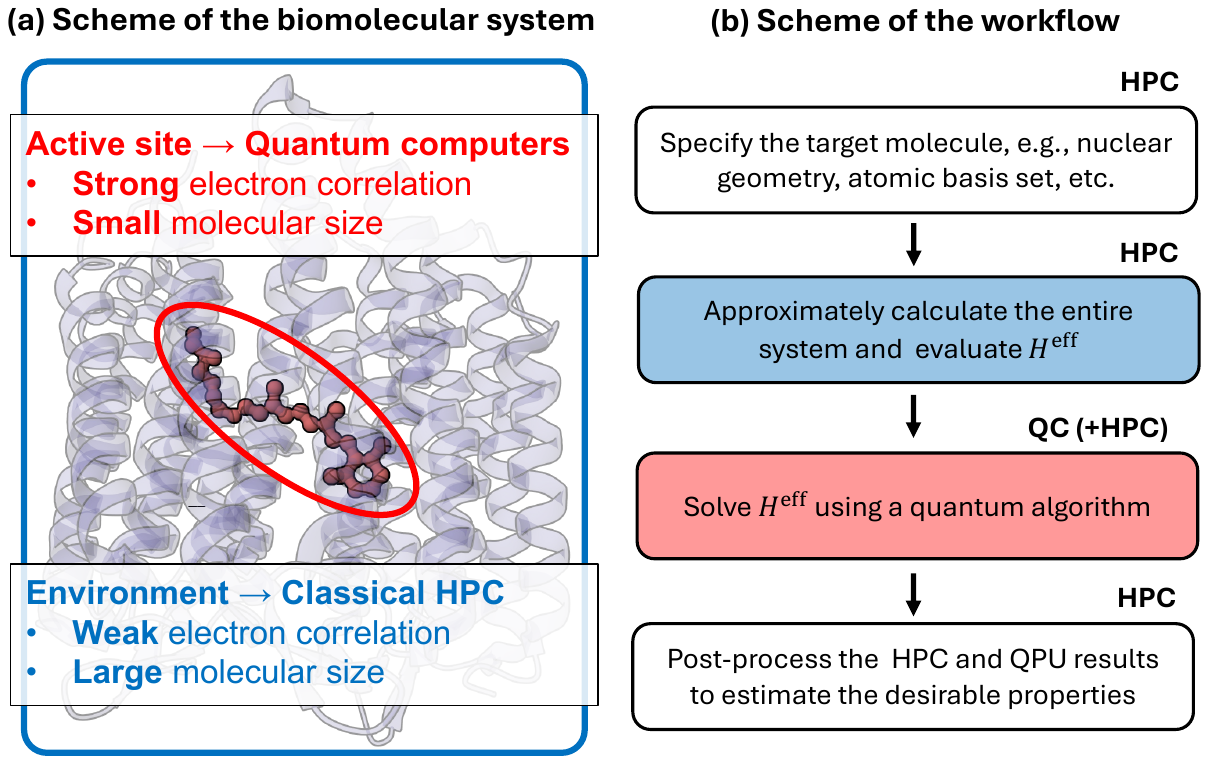}
    \caption{
        (a) Schematic representation of the biomolecular system consisting of the active site and the environment.
        (b) Typical scheme of the workflow. The active site is treated with quantum computing (QC), while everything else, including the computationally intensive environment, is handled on HPC. This layered approach enables efficient, accurate simulations of complex biomolecular reactions.
    }
    \label{fig:system}
\end{figure}

Layered approaches such as quantum mechanics/molecular mechanics (QM/MM)~\cite{Warshel1976-ua} and ONIOM\footnote{ONIOM = Our own N-layered Integrated molecular Orbital and Molecular mechanics} methods~\cite{Vreven2006-sq} have been developed to address these challenges by partitioning the system into two or more parts, each treated by a different level of theory (see Fig. \ref{fig:system}(a)). The MM part typically employs scalable approximate methods in molecular size, such as force fields, to model the large environment~\cite{Karplus1990-wd}, while the QM part focuses on the smaller active site with higher accuracy, using methods such as first-principles quantum chemistry calculations~\cite{Helgaker2014-ce}.

The practical bottleneck to overall accuracy lies on the QM side, where strong electron correlation often occurs. Proper representation of these interactions for an accurate description of the chemical reactions is particularly challenging with existing approximate QM methods such as density functional theory (DFT). Quantum computing~\cite{Nielsen2010-ki} has emerged as a promising avenue for accelerating the simulation of strongly correlated systems, owing to the known exponential quantum speedup for certain algorithms, such as quantum phase estimation (QPE)~\cite{Nielsen2010-ki}. Thus, a hybrid quantum-HPC approach naturally fits the layered methods, where HPC stands for (classical) high-performance computing. A typical scheme of the workflow is shown in Fig. \ref{fig:system}(b).

To aim for quantum advantage in this field before a fault-tolerant quantum computer becomes available, we select a promising combination of target molecules and methods. Herein we propose photo-induced biomolecular chemical reactions~\cite{Curutchet2017-ek, Hampp2000-eu, Acharya2017-zt} and quantum selected configuration interaction (QSCI) ~\cite{Kanno2023-dn, Mikkelsen2024-la, Kenji2024-xa, Sugisaki2025-zj, Erhart2025-bf, Robledo-Moreno2025-zq, Shirakawa2025-vt, Shajan2025-wt} as one such promising combination. Chemical reactions often involve transition states with a drastic change in electronic states, for which conventional approaches often struggle to accurately estimate chemical properties. QSCI is a selected CI method~\cite{Harrison1991-zu, Holmes2016-kq} that uses quantum circuits to sample relevant electronic configurations.
We aim to augment the conventional wavefunction-based approach on HPC with a quantum computer to achieve a more accurate estimate of chemical properties than those of HPC alone within practical computing time.

In this letter, we develop a method based on time-evolved QSCI (TE-QSCI)~\cite{Mikkelsen2024-la, Kenji2024-xa, Sugisaki2025-zj} to efficiently extend the configuration space previously constructed on HPC, and implement an end-to-end workflow on a large-scale hybrid computing system to integrate quantum computing into the layered approach.
To focus on the methodological aspects, we use simple wavefunction-based methods and model systems that qualitatively represent the characteristics of biomolecular systems.
An accurate description of the real biomolecular system is not in the scope of the present work.
We show that the present method can efficiently find additional configurations in comparison to classical approaches.

Below is the overview of the present demonstration. We apply the ONIOM method~\cite{Vreven2006-sq} as a layered framework, and use the TE-QSCI-based method for the active site to capture additional electron-correlation effects beyond those preliminarily estimated by HPC. The environment is managed using the conventional Hartree--Fock or configuration interaction singles (CIS) method~\cite{Helgaker2014-ce}, enabling us to leverage standard computational resources and implement a workflow across the hybrid computing system. We take a workflow-based approach to hybrid computation to facilitate the use of existing HPC resources.

We use the hybrid system of supercomputer Fugaku and the Quantinuum Reimei as the computational platform.
To maximize the efficiency of the hybrid workflow, it is integrated with a dedicated network and job-scheduling system. While the HPC job is submitted to Fugaku via the existing job scheduler, the quantum job is submitted to Reimei via the direct-access API, rather than through a typical cloud interface.
We use 96 Fugaku compute nodes (A64FX) for all HPC calculations, including geometry optimization and baseline electronic structure calculations. Quantum computations are performed on Reimei, a 20-qubit system with all-to-all connectivity. Hybrid quantum-HPC computations are orchestrated using the Tierkreis~\cite{Sivarajah2022-ik} workflow system, which facilitates the coordination between the classical and quantum computational tasks.
Quantum circuits are built and optimized using TKET~\cite{Sivarajah2020-jg}.

\textit{Model systems}.
We take the photoisomerization of a retinal molecule as a typical example, expressed as
\begin{align}
    \includegraphics[width=0.99\hsize]{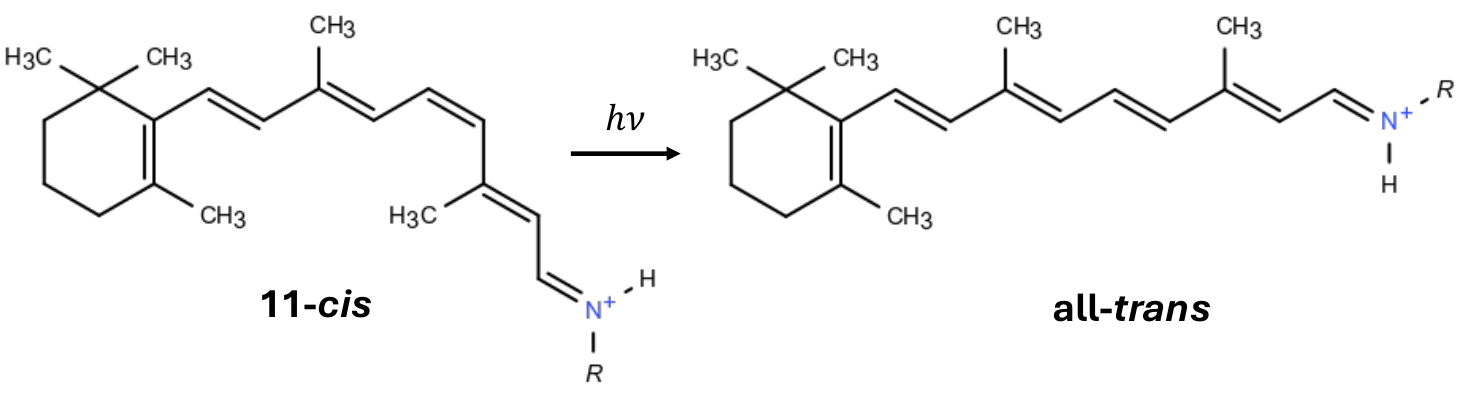}
\end{align}
where \textit{R} represents a model environment system.
The photoisomerization of retinal underlies many of the photo-induced chemical reactions in biological systems, including photo-induced ion pumping~\cite{Hampp2000-eu}. 
This reaction has a significant multiconfigurational character~\cite{Gonzalez-Luque2000-zx}.

\begin{figure}
    \centering
    \includegraphics[width=0.99\hsize]{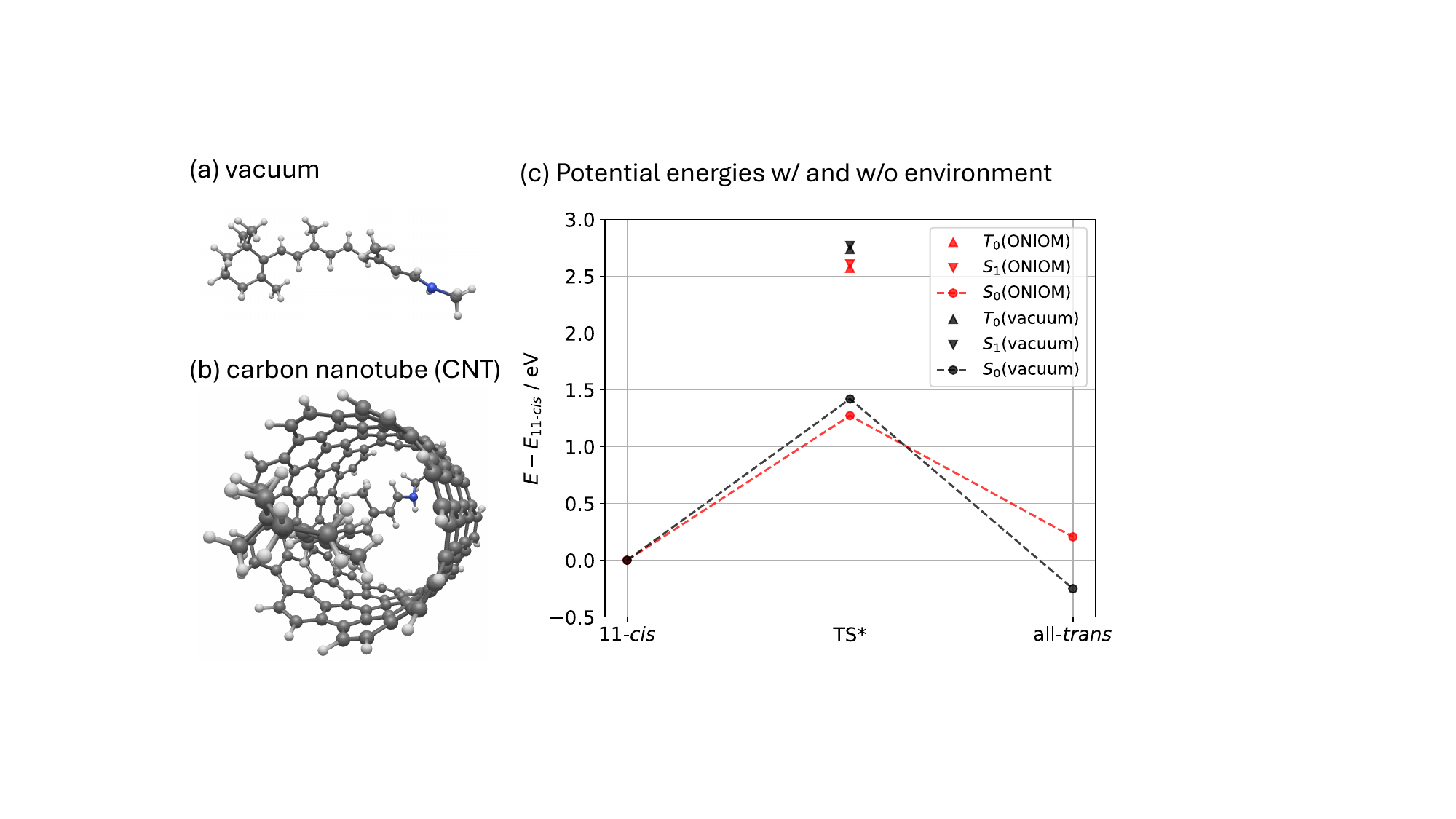}
    \caption{
        (a) Model system in vacume and (b) model system embedded CNT, both in the TS* state for biomolecular excited states involving conical intersections. The retinal molecule serves as the active site, while the CNT provides the environmental effect.
        (c) Potential energies relative to the energy of 11-\textit{cis} for each model system. We connect the ground-state energies to facilitate comparison.
    }
    \label{fig:retinal_cnt}
\end{figure}

To simplify the model system, we use a carbon nanotube (CNT) as the surrounding environment for the retinal~\cite{Liu2007-je}. CNT serves as a ``wall'' to provide a steric effect to the retinal covalently bonded to it on its edge.
It may alter the relative energies of the (meta)stable isomers and transition states, thereby influencing the selectivity and kinetics of chemical reactions. We add H atoms to terminate the dangling bonds of CNT. We consider $R$ = CH$_{3}$ (Fig. \ref{fig:retinal_cnt}(a)) to model the isolated system and $R$ = (CH$_{2}$)$_{2}$--CNT to model the system embedded in an environment (Fig. \ref{fig:retinal_cnt}(b)).

\textit{ONIOM method}.
The basic idea is to evaluate the total energy $E_{\text{ONIOM}}$ by breaking the molecular system into two layers and applying different levels of theory to balance the overall accuracy and the scalability. In this scheme, $E_{\text{ONIOM}}$  is represented as
\begin{equation}
    E_{\text{ONIOM}} = E_{\text{low}}^{\text{real}} + E_{\text{high}}^{\text{model}} - E_{\text{low}}^{\text{model}},
\end{equation}
where $E_{\text{low}}^{\text{real}}$ is the energy of the entire system calculated using a low-level method and $E_{\text{high}}^{\text{model}}$ is the energy of the active site calculated using a high-level method. The ONIOM method is often called a subtractive scheme, as the last term, $E_{\text{low}}^{\text{model}}$, is subtracted to avoid double-counting the active-site energy evaluated with the low-level method.

Here, we use the QM1/QM2 ONIOM flavor, in which both the high- and low-level parts are treated quantum-chemically. We focus on the singlet ground state $S_{0}$, the first excited state ($S_{1}$) and the triplet ground state $T_{0}$. We use the CIS/RHF/6-31G(d,p) level of theory for the low-level method, and complete active space configuration interaction (CASCI) on top as the high-level method with an active space of 8 electrons in 8 spatial orbitals, represented as CASCI(8,8). The 11-\textit{cis} and all-\textit{trans} geometries are obtained at the RHF/6-31G level. The geometries of $S_{1}$/$T_{0}$ conical section (referred to as TS*) were heuristically obtained by optimizing the geometry of $S_{1}$ state at the CIS/RHF/6-31G level. The total energy relative to $S_{0}$ of the 11-\textit{cis} geometry is shown in Fig.~\ref{fig:retinal_cnt}(c). We use the \texttt{NTChem} software package~\cite{Nakajima2015-rp} for all the classical computational chemistry calculations.

As shown in Fig. \ref{fig:retinal_cnt}(c), the energy profile obtained with the ONIOM method is altered by the CNT environment due to the steric effect.
The 11-\textit{cis} isomer is relatively stabilized by the CNT environment, as the all-\textit{trans} isomer suffers more from steric hindrance with the CNT wall in this particular case. Whereas the energy of the TS* geometry is also stabilized by the CNT environment, the $S_{0}\to S_{1}$ excitation energy is not very different from that without the environment. We attribute this behaviour to the presence of similar steric effects in both the $S_{0}$ and $S_{1}$ states at the TS* geometry.

\textit{TE-QSCI}.
To demonstrate the computation of $E_{\text{high}}^{\text{model}}$ using quantum computer, we proceed to use the TE-QSCI method. In QSCI~\cite{Kanno2023-dn}, we collect the electronic configuration samples $\mathcal{S}$ using an ansatz quantum circuit to construct a Krylov subspace Hamiltonian matrix,
\begin{align}
    H^{\text{QSCI}}
    =
    P^{\dagger}
    H
    P
    ,
    ~~
    P = \sum_{x\in \mathcal{S}} \ket{x}\bra{x}
    ,
\end{align}
followed by its diagonalization on HPC to evaluate approximate ground (and excited) states. Whereas the original proposal~\cite {Kanno2023-dn} employed a suboptimal parameterized quantum circuit (ansatz), TE-QSCI uses time evolution to sample configurations from the Krylov subspace. 

We extend TE-QSCI, so that it improves the description of the electron correlation from the classical baseline to reliably leverage the HPC resources. Instead of taking the single-reference (e.g., Hartree--Fock) state as an initial configuration, we approximately input the CASCI state for a smaller active space, and then let it evolve in time. The present TE-QSCI ansatz is expressed as
\begin{align}
    \footnotesize
    \begin{quantikz}[style={column sep=5pt, row sep=8pt}]
        \lstick{$q_{0}: \ket{0}$} & \gate{X} & \gate[7]{\e^{-\im(H-H^{0})t}} & \meter{}
        \\
        \lstick{$q_{1}: \ket{0}$} & \gate{X} & & \meter{}
        \\
        \lstick{$q_{2}: \ket{0}$} & \gate[3]{V_{j}^{0}} & & \meter{}
        \\
        \lstick{$\vdots$} 
        \\
        \lstick{$q_{13}: \ket{0}$} & & & \meter{}
        \\
        \lstick{$q_{14}: \ket{0}$} & & & \meter{}
        \\
        \lstick{$q_{15}: \ket{0}$} & & & \meter{}
    \end{quantikz}
\end{align}
The present TE-QSCI method begins with a state preparation that encodes an approximate wavefunction $\ket{\Phi_{j}^{0}}$, satisfying
\begin{align}
    H^{0}\ket{\Phi_{j}^{0}}
    =
    E^{0}_{j}\ket{\Phi_{j}^{0}}
\end{align}
where $E_{j}^{0}$ and $\ket{\Phi_{j}^{0}}$ are the $j$-th eigenvalue and eigenstate of $H^{0}$, respectively. $H^{0}$ is the Hamiltonian defined in the initial active space. $t \in \mathbb{R}$ is a parameter representing the time-evolution period.
We use the Jordan--Wigner transformation to map the electronic Hamiltonian to the qubit counterpart. We then perform time evolution under the operator $\e^{-\im (H-H^{0})t}$ to explore the configuration space and identify configurations that contribute to the ground and low-lying excited states.
We can reduce the number of terms using $H - H^{0}$ to obtain essentially the same state, because we input the eigenstate of $H^{0}$ with $V_{j}^{0}$.
We merge the samples from all the time steps~\cite{Mikkelsen2024-la} and construct a subspace Hamiltonian matrix, from which we compute $E_{\text{high}}^{\text{model}}$.

We use a (6,6) active space followed by time evolution with the (8,8) active space. 
$V_{j}^{0}$ is prepared using adaptive variational algorithm~\cite{Grimsley2019-yc} to maximize $|\braket{\Phi_{j}^{0}|V_{j}^{0}|0}|^{2}$ with qubit excitatitation operators~\cite{Yordanov2022-qj}.
We sample the circuit with all the pairs of $\Delta t \in \{10^{-3}, 2.5, 5.0, 7.5\}$ in atomic unit and $j = 0, 1, 2$ corresponding to $S_{0}$, $S_{1}$, and $T_{0}$, respectively. 1500 shots are taken for each pair (i.e., 18000 shots in total). The time-evolution circuit is implemented with two Trotter steps, i.e., $t = 2\Delta t$. The time-evolution circuit is truncated to include up to 500 two-qubit gates per Trotter step of $\Delta t$. We post-select the measurement outcomes satisfying the particle and spin conservation ($S_{z}$ projection). In addition, we infer that the configurations include those with the same spatial orbital occupations. For example, if we identify a configuration with two singly occupied spatial orbitals, we add two configurations with
$\cdots\uparrow\downarrow\cdots$ and $\cdots\downarrow\uparrow\cdots$ occupation patterns.
In this work, leveraging all-to-all connectivity, we demonstrate that ``important'' configurations beyond the classical baseline can be efficiently identified with the time-evolution circuits for a 16 logical qubit system. Demonstration with a larger number of qubits is scheduled for future work.

\textit{Logical circuit encoding}.
Logical quantum circuits are encoded by the $\llbracket n, k, d\rrbracket$ error detection code with $(n, k, d) = (18, 16, 2)$ (dubbed iceberg code)~\cite{Self2024-vo}, where $n$, $k$, and $d$ denote the number of physical qubits, the number of logical qubits, and the code distance, respectively. We detect a weight-one error, and the measurement outcomes are postselected. See Refs.~\cite{Self2024-vo, Yamamoto2024-vm, Jin2025-rt} for its applications.


Hereafter, we focus on the results of the TE-QSCI calculations for the active site (retinal molecule), i.e., $E_{\text{high}}^{\text{model}}$ of the TS* geometry, since the ONIOM workflow is identical to the conventional one. Table~\ref{tab:excitation_energy} summarizes the excitation energies obtained from different methods, Hartree--Fock (HF), CASCI(2,2),  CASCI(6,6) (classical baseline), TE-QSCI(8,8) using Reimei, and CASCI(8,8). CASCI(6,6) and CASCI(8,8) have 400 and 4900 configurations, respectively. 

\begin{figure}
    \centering
    \includegraphics[width=0.89\hsize]{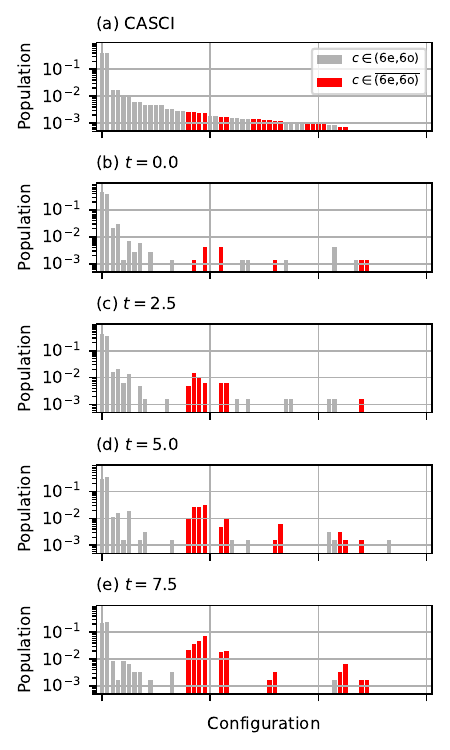}
    \caption{
        Histogram of the measurement outcomes $c$ (configurations) for the S$_1$ state of retinal embedded in a carbon nanotube. The horizontal axis represents the configuration indices, while the vertical axis indicates the probability $p$ of each configuration being measured. The red bars represent the configurations to be identified by TE-QSCI, while the gray bars indicate those identified in the classical baseline calculation. (a) exact CASCI and (b)-(e) TE-QSCI results from Reimei with different time step sizes.
    }
    \label{fig:histogram1}
\end{figure}

Our task is to identify the important configurations outside this initial active space using TE-QSCI. Fig.~\ref{fig:histogram1} shows the (normalized) histogram of the measurement outcomes for the $S_1$ state from both the exact CASCI and TE-QSCI calculations.
TE-QSCI identifies 2149 ``new'' configurations from 12214 shots. The average survival rate of the iceberg code is 67.9\%.
The TE-QSCI identifies the first 6 configurations in the CASCI(8,8) calculations with high probability, suggesting that the present method efficiently captures the essential electron-correlation effects. Those for the $T_{0}$ state show a similar behavior, while the S$_{0}$ state shows less improvement from the classical baseline, as the molecular orbital basis is optimized for $S_{0}$ through the HF method (See Appendix \ref{app:histogram}).

It is remarkable that we experimentally observe that the longer time evolution enhances the probability of measuring ``new'' configurations (See Fig. \ref{fig:histogram1}(b)-(e)).
This is interpreted considering the Krylov subspace.
The time-evolution is rewritten as
\begin{equation}
    \ket{\Psi(t)} = \e^{-\im H t}\ket{\Phi_{0}} = (1 - \im Ht)\ket{\Phi_{0}} + O(t^{2})
\end{equation}
This expression suggests that TE-QSCI identifies higher-order Krylov basis regarding $H$ with higher probability, as $t$ increases. Given the quantum simulation exhibits quantum speedup~\cite{Abrams1997-xz}, TE-QSCI is promising in complementing the classical methods with HPC alone.

Energies and fidelity of the approximate eigenstates to CASCI(8,8)  using different methods are summarized in Table~\ref{tab:excitation_energy}.
\begin{table}[htbp]
    \centering
    \caption{
        Energy and fidelity comparisons between different approximate methods.
        Correlation energy $E_{\text{corr}} = E_{0} - E_{\text{HF}}$ with $E_{0}$ and $E_{\text{HF}}$ denoting $S_{0}$ and the HF energies,
        excitation energy $\Delta E = E_{j} - E_{0}$,
        and
        fidelity $f = |\braket{\Phi|\text{CASCI(8,8)}}|^{2}$ with $\ket{\Phi}$ representing the eigenstate obtained for each method.
    }
    \begin{tabular}{c|r|rr|rrr}
        \hline\hline
                     & $E_{\text{corr}}$/eV & \multicolumn{2}{c}{$\Delta E$}/eV & \multicolumn{3}{|c}{$f$} \\
        \hline 
                     &            & $S_{1}$           & $T_{0}$          & $S_{0}$       & $S_{1}$      & $T_{0}$      \\
        \hline 
        HF           & $ 0.00 $    &  $ ~-~  $        &  $  ~-~   $     &  $ 0.940 $   &  $ ~-~  $   &  $ ~-~   $  \\
        CASCI(2,2)   & $ 0.00 $    &  $ 2.24 $        &  $  2.28  $     &  $ 0.940 $   &  $ 0.831$   &  $ 0.835 $  \\
        CASCI(6,6)   & $-0.43 $    &  $ 1.44 $        &  $  1.48  $     &  $ 0.975 $   &  $ 0.955$   &  $ 0.954 $  \\
        TE-QSCI(8,8) & $-0.69 $    &  $ 1.23 $        &  $  1.27  $     &  $ 0.994 $   &  $ 0.995$   &  $ 0.996 $  \\
        CASCI(8,8)   & $-0.80 $    &  $ 1.26 $        &  $  1.29  $     &  $ 1.000 $   &  $ 1.000$   &  $ 1.000 $  \\
        \hline  \hline 
    \end{tabular}
    \label{tab:excitation_energy}
\end{table}
Recall that the excited states show greater fidelity improvement.

The TE-QSCI results show a significant improvement over the initial classical baseline, both in excitation energy and state fidelity. The TE-QSCI(8,8) results show good agreement with the exact CASCI(8,8) calculations despite the modest number of shots ($\sim 10^4$).
Exploring the practical scaling of the number of shots with the present TE-QSCI is saved for future work.

Another aspect of the scalability is the circuit complexity. Naively, the current workflow is unlikely to scale beyond the limitation of brute-force simulation on HPC with $\sim 50$ qubits, as the time-evolution circuit scales at $O(N^{4})$, where $N$ is the number of qubits. Further development of approximate methods to reduce the circuit size will be required. Preliminary circuit optimization on HPC, such as in methods using tensor networks~\cite{Kenji2024-xa}, may be an option. 


In conclusion, we have developed a quantum-HPC hybrid workflow for biomolecular ground and excited-state calculations and have demonstrated it on a hybrid computing system comprising the supercomputer Fugaku and the quantum computer Reimei. We extended time-evolved quantum selected configuration interaction (TE-QSCI) to enable a quantum computer to efficiently identify configurations outside the active space that are within the reach of HPC alone. This framework can readily be adapted to other layered approaches, such as QM/MM. Our results highlight the potential of quantum computing to enhance the accuracy of biomolecular simulations, particularly for chemical reactions involving excited states.

In the long term, the quantum algorithm will be reconsidered as the quantum computer capabilities advance. Notably, the QPE algorithm provides an exponential quantum speedup. The present approach shares common primitives with the QPE algorithm, namely time-evolution and state-preparation circuits. We hope this work paves the way for scalable and accurate simulations of complex biomolecular reactions using a quantum-HPC hybrid approach.


\textbf{Data availability}. Data presented here is accessible upon request.

\textbf{Acknowledgments}.
We are grateful to the entire Quantinuum team for their many contributions to the work.
We thank
Kengo~Nakajima,
Shinji~Sumimoto,
Yuetsu~Kodama,
Kazuya~Yamasaki,
Tamiya~Onodera,
Duncan~Gowland,
Luca~Erhart,
Grahame~Vittorini,
for useful discussions and feedback on the manuscript.
Part of this work is based on results obtained from a project, JPNP20017, commissioned by the New Energy and Industrial Technology Development Organization (NEDO).
This work used computational resources of Fugaku provided by RIKEN Center for Computational Science (Project ID: ra010014).

\bibliography{main}
\onecolumngrid
\clearpage
\appendix
\renewcommand\thefigure{\thesection \arabic{figure}}
\renewcommand\thetable{\thesection \arabic{table}} 
\onecolumngrid
\section{Histogram of the measurement outcomes\label{app:histogram}}

\begin{figure}[b]
    \centering
    \includegraphics[width=0.89\hsize]{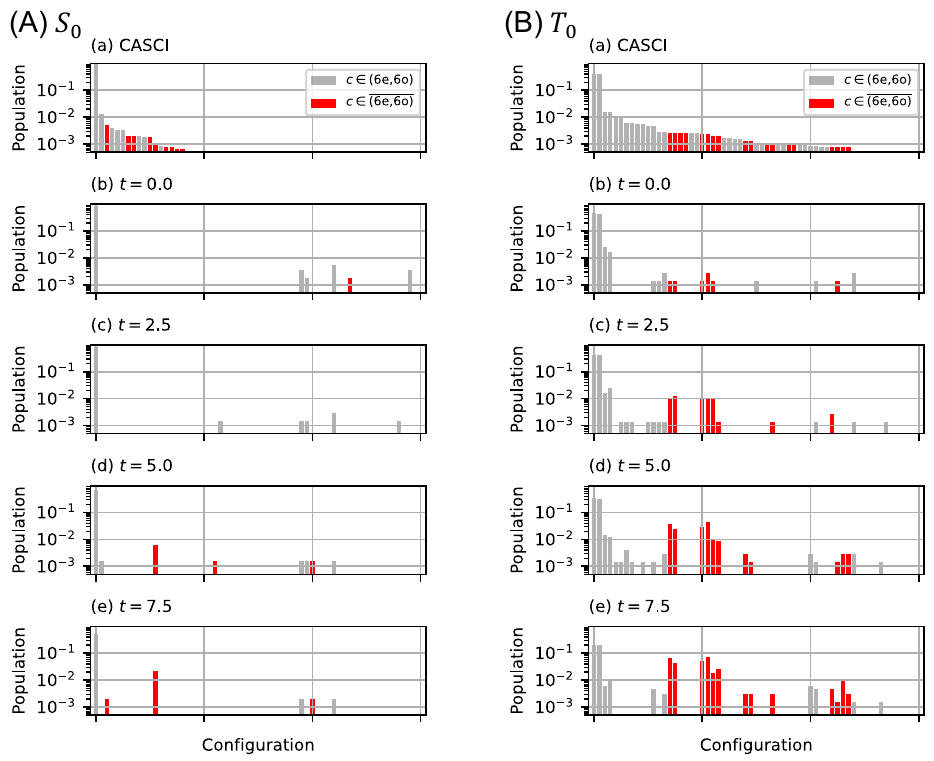} 
    \caption{
        Histogram of the measurement outcomes $c$ (configurations) for (A) S$_0$ and (B) $T_{0}$ states. For each panel, (a) exact CASCI and (b)-(e) TE-QSCI results from Reimei with different time step sizes.
        \label{fig:histogram_s0_t1}
    }
\end{figure}

The histogram for $S_{0}$ and $T_{0}$ are shown in Fig. \ref{fig:histogram_s0_t1}.
Whereas $T_{0}$ shows a similar behavior to $S_{1}$ shown in Fig. \ref{fig:histogram1}, $S_{0}$ finds a modest number of ``new'' configurations, as the initial state has significantly higher fidelity.

\end{document}